# Polarization analysis and Density Functional Theory (DFT) simulations of the electric field in InN/GaN multiple quantum wells (MQWs)

Z. Romanowski[1], P. Kempisty[1], K. Sakowski[1] and S. Krukowski[1,2]

[1]Institute of High Pressure Physics, Polish Academy of Sciences, Sokołowska 29/37, 01-142 Warsaw, Poland

[3]Interdisciplinary Centre for Materials Modelling, Warsaw University, Pawińskiego 5a, 02-106 Warsaw, Poland

## Abstract

Results of the first ab initio simulations of InN/GaN multiquantum well (MQW) system are presented. The DFT results confirm the presence of the polarization charge at InN/GaN interfaces, i.e. at polar InN/GaN heterostructures. These results show the potential jumps which is related to the presence of dipole layer at these interfaces. An electrostatic polarization analysis shows that the energy minimum condition can be used to obtain the field in InN/GaN system, employing standard polarization parameters. DFT results are in good agreement with polarization data confirming the existence of electric field leading to separation of electron and holes in QWs and emergence of Quantum Confined Stark Effect (QCSE).

Gallium, aluminum and indium nitrides are presently considered as principal materials for constructing quantum well (QW)based optoelectronic devices: laser diodes (LDs) and light emitting diodes (LEDs) [1]. In spite of considerable progress in physical understanding and technological implementation of these materials, some fundamental problems remain to be solved to open routes for further progress. Since overwhelming majority of these devices is constructed on polar GaN (0001) surface, natural consequence of such construction is presence of built-in and strain induced electric fields [2,3]. The electric field changes the energy levels of both type of carriers, electrons and holes, the phenomenon known as Quantum Confined Stark Effect (QCSE) [5,6]. Yet far more sinister consequences are related to the fact that the electrons and holes are shuffled to the opposite ends of the quantum well [7,8]. This effect significantly reduces the overlap of their wavefunctions, and accordingly lowers radiative recombination rates, and finally, the efficiency of optoelectronic devices, both LDs and LEDs [9-12]. The negative influence of QCSE effect may be enhanced by Auger recombination or carrier leakage for high carrier density at high injection currents [13-16]. That could lead to decrease of the device efficiency for higher injection currents, nicknamed as "efficiency droop" [17]. Even in typical blue LD and LEDs, based on 10 at% indium QWs, the QCSE is supposed to cause a considerable separation of the carriers and dramatic reduction of the efficiency of the devices. Even more drastic influence is expected to exist in more In-rich green LEDs and LDs which are currently developed. Therefore the QCSE was studied intensively, using both theoretical and experimental methods. The experimental research was directed mostly toward investigation of size dependence of the emission from MQW structures [6,7]. It is commonly assumed that a fingerprint of the presence of the field separation of the carriers is redshift of the emission line and dramatic increase of the lifetime for wider wells [9,10]. A second possible indication of the QCFE is the presence of the screening effect, which can be controlled by the doping of the barriers of the QW system [18]. Typically, n-type doping, by Si donors is used for that purpose. Another possible approach would be to use polarization-matched GaInN/AlGaN MQW in order to reduce carrier leakage to remove the efficiency droop altogether [19]. It is understandable that the planning of these approaches is based on simplified treatment of the problem and identification of the main trends in the system. Since the electro-optical emitting MQW system in both LEDs and LDs is located in the vicinity of the p-n junction, the presence of additional field can, in principle, affect the properties of the system seriously, and change the observed trends in the optically emitting system. Therefore an interesting solution is to use optical pumping of the wells, using limited or high optical power. This could help to identify the presence of screening effects in the observed spectra. Among other methods that can be used to investigate the piezoelectric effects, the pressure investigations play important role [19]. Another approach is to grow MQW on semipolar planes which could potentially remove the piezoelectric contributions completely [20]. Nevertheless the progress in understanding of these systems is slow because the researchers need to cope with additional, potentially strong effects which could affect the results, such as the presence of the extended defects within the well system or the segregation of indium. The latter could, in principle, transform 2-d translationally invariant QW system into a collection of quantum dots, having drastically different symmetry and consequently the electrical and optical properties. Therefore interpretation of the results of the MQWs measurements is fairly nontrivial task and additional insight from the theoretical analysis could be very helpful. The theoretical approaches were used in order to shed some light onto various aspect of the problem. Since the QW system was considered difficult to tackle by ab initio methods due to its shear size, the approximate step by step methods were developed. A most systematic approach along this line is to obtain the spontaneous polarization and piezoelectric tensor of bulk material, i.e. GaN or InN which, via the deformation tensor can be used to determine the magnitude of polarization in the strained thin layer. Such approach is typically used in estimation of the magnitude of the effect in mixed GaInN/GaN QWs. Here we adopt a different route by direct simulation of the electric properties of GaN/InN QW systems. The InN/GaN MQWs are more difficult technologically than standard GaN/InN MQW system. A considerable technological advances were made which make possible investigation of the InN/GaN MQW systems by both electrical and optical measurements [21,22]. The InN/GaN are still of relatively poor quality, but their construction opens the route to potential investigation of these systems. As InN/GaN MQWs are relatively easy to be tackled by ab initio simulations, they offer a possibility of comparison of experimental data and DFT simulations.

The freely accessible DFT SIESTA code, used in our calculations, combines norm conserving pseudopotentials with the local basis functions [23-25]. The pseudopotentials for Ga, H and N atoms were generated, using ATOM program for all-electron calculations [26,27]. Gallium 3d electrons were explicitly included in the valence electron set. For Ga, H and N atoms the double $\zeta$ local basis set was used with polarization. LDA CA and basis set with soft confinement potential were used in the tests, with the following values for the lattice constants of bulk crystals: GaN - a = 3.181 Å and c = 5.146 Å, and InN - a = 3.527 Å and c = 5.713 Å. These values are in good agreement with the experimental data for GaN: a = 3.189 Å and c = 5.185 Å and for InN: a = 3.52 Å and c = 5.72 Å. The results reported below were obtained using LDA CA approximation and basis orbitals generated with soft confinement potential. All additional data on the simulation procedure may be found in Ref. 29 [29].

In typical LED or LDs, the MQW system can be located in the electric field created by p-n junction or outside this area [17]. In the first case, the electric conditions change depending on the electric field applied to the device. In the second, the electric potential distribution is universal. The potential distributions presented in Fig 1 reflects the fact that the Fermi energy is controlled by dominant point defect in the bulk of semiconductor, i.e. it is located at the same energy level at both sides of MQW system. This case is considered in the present paper.

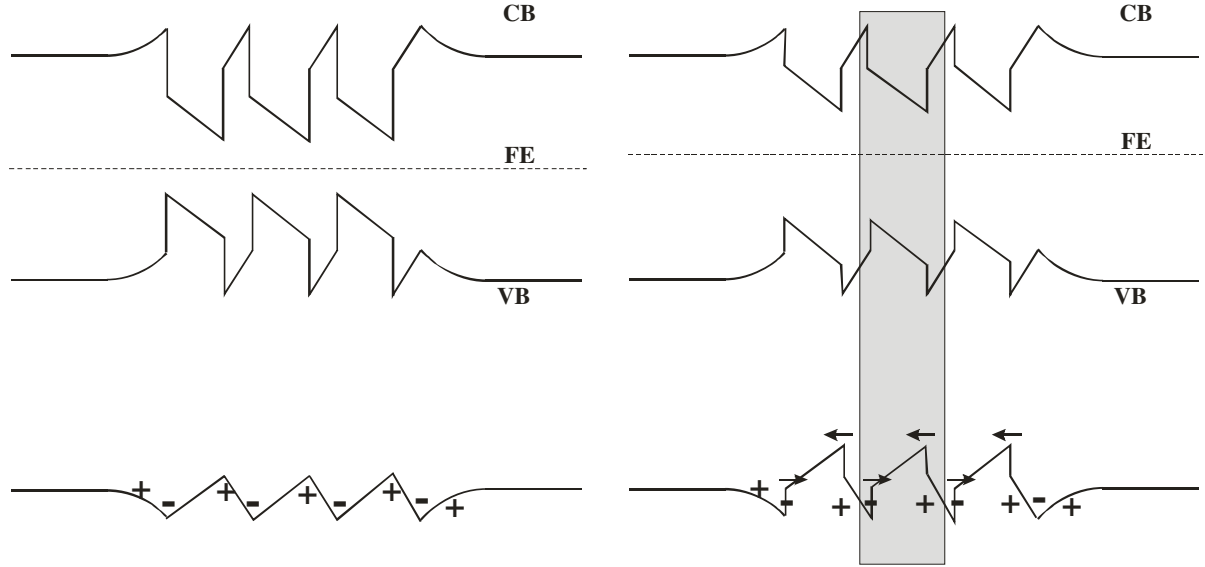

*Fig.1 MQW system outside p-n junction area with band structure and the distribution of electric potential and the polarization and screening charges: left – standard picture with polarization charges only; right – DFT picture with polarization charges and the dipole layer at the interfaces. Shaded - area modeled in DFT simulations.*

Accordingly, the electric potential change across the entire MQW system, located far from p-n junction, is zero. That leads to cancelling contribution of polarization and screening charges as shown in Fig 1. The standard Periodic Boundary Conditions (PBC) of the simulated area, denoted by gray in Fig. 1, can be adopted for electric potential. PBC is currently used in large majority of DFT simulations. Recently, an efficient Laplace correction scheme was presented which allows us to alleviate this condition preserving efficiency of the calculations and stability of self consistent field (SCF) iterative solution scheme [30]. Similarly PBC was used in the DFT investigations of AlN/GaN superlattices [31]. This is in accordance to standard assumption of zero far distance field in the solids, also for insulators [32]. It is argued that the macroscopic systems attain the energy minimum, characterized by zero electric field, the condition which, in the case of insulators, can take even several hours to attain [33,34]. This assumption is a foundation of Berry phase approach, used in determination of polarization of insulating solids [33-35]. The simulated (1 x 1) supercell consists of InN well and the GaN barrier of the thickness $z_{InN}$ and $z_{GaN}$, respectively, which imposes following bound for z-components of the electric fields in InN and GaN layers:

$$E_{3,GaN}\, z_{GaN} + E_{3,InN}\, z_{InN} + \Delta V_1 + \Delta V_2 = 0 \qquad (1)$$

where, as shown below, $\Delta V_1$ and $\Delta V_2$ are the potential jumps due to existence of dipole layer in the GaN-InN and InN-GaN polar heterostructes. These dipole layers have not been employed in the description of nitride systems. Eq.1. can be used to remove the electric field in the barrier i.e. $E_{3,GaN}$ from the energy minimization procedure. Typically, the screening field in the semiconductor interior compensates the field, created by spontaneous polarization. In the case when the layer is much thinner than the screening length, the field is different: the long distance field is absent, and the minimal energy state is attained for the uniform field created by spontaneous polarization and piezoelectric effects. Still in MQWs the uniform field is incompatible PCB conditions given by Eq. 1, and accordingly the field arising in the structure should attain the minimum energy state, i.e. minimizing the excess energy functional $\Delta f$:

$$\Delta f = \frac{1}{2\varepsilon_0} \int d^2r \left[ \int_0^{z_{InN}} dz \frac{\left(E_{3,InN} - E_{3,InN}^{(0)}\right)^2}{\varepsilon_{33,InN}^\infty} + \int_0^{z_{GaN}} dz \frac{\left(E_{3,GaN} - E_{3,InN}^{(0)}\right)^2}{\varepsilon_{33,InN}^\infty} \right] \qquad (2)$$

where $E_{3,InN}^{(0)}$ and $E_{3,GaN}^{(0)}$ are the fields arising from the spontaneous and strain induced polarization in strained bulk InN and GaN [2,4], given by:

$$E_3^{(0)} = \frac{D_3^{(0)}}{\varepsilon_0 \varepsilon_{33}^\infty} = \frac{P_3}{\varepsilon_0 \varepsilon_{33}^\infty} = \frac{\left[P^{eq} + e_{33}\in_3 + 2e_{31}\in_1\right]}{\varepsilon_0 \varepsilon_{33}^\infty} \quad (3)$$

where index denoting material was omitted for simplicity (notation of Bernardini et al. was used [3]). The InN and GaN spontaneous polarization and piezoelectric constants used, were derived from Ref 3. For GaN these values are: $P_{GaN}^{eq} = -0.029 C/m^2$, and $e_{33,GaN} = 0.73 C/m^2$, $e_{31,GaN} = -0.49 C/m^2$, and for InN: $P_{InN}^{eq} = -0.032 C/m^2$, and $e_{33,InN} = 0.97 C/m^2$, $e_{31,InN} = -0.57 C/m^2$. The strain values $\in_3$ and $\in_1$ (Voigt notation) were derived from the average positions of atoms in the supercell after relaxation. Employing minimum condition to the excess energy functional (3), subject to constraint (1), the following electric fields were obtained: .

- in InN well:

$$E_{3,InN} = \frac{\left(\varepsilon_{InN} E_{3,InN}^{(0)} - \varepsilon_{GaN} E_{3,GaN}^{(0)}\right) z_{GaN} - \varepsilon_{GaN}(\Delta V_1 + \Delta V_2)}{\left(\varepsilon_{InN} z_{GaN} + \varepsilon_{GaN} z_{InN}\right)} \quad (4a)$$

- in GaN barrier:

$$E_{3,GaN} = \frac{\left(\varepsilon_{GaN} E_{3,GaN}^{(0)} - \varepsilon_{InN} E_{3,InN}^{(0)}\right) z_{InN} - \varepsilon_{InN}(\Delta V_1 + \Delta V_2)}{\left(\varepsilon_{InN} z_{GaN} + \varepsilon_{GaN} z_{InN}\right)} \quad (4b)$$

The potential jumps $\Delta V_1$ and $\Delta V_2$ have to be obtained from DFT simulations.

The DFT results, presented in Fig.2, were obtained for constant barrier width of sixteen Ga-N atomic layers (ALs), i.e. it is approximately 2.5 nm thick. The InN quantum well consisting of four, six and eight In-N ALs were simulated. The diagrams show the potential profiles within the InN – GaN system.

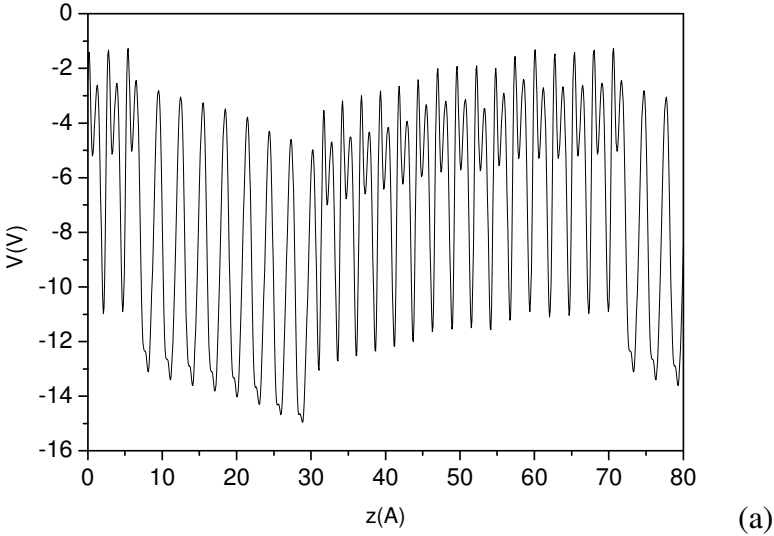

(a)

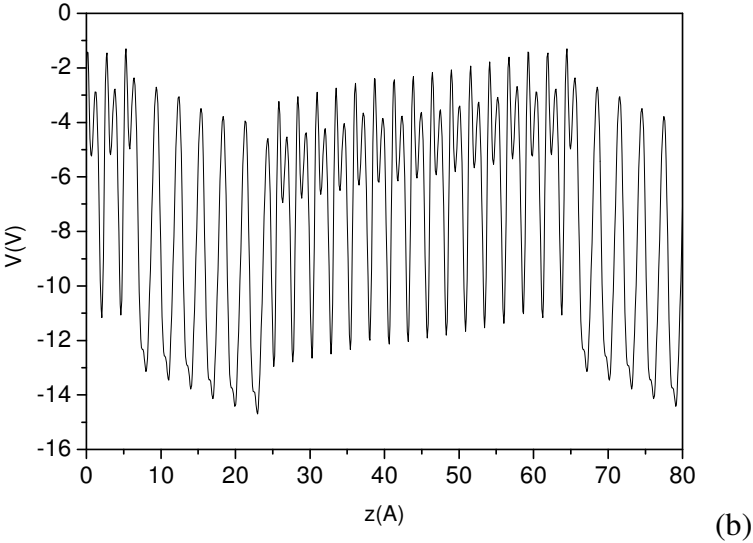

(b)

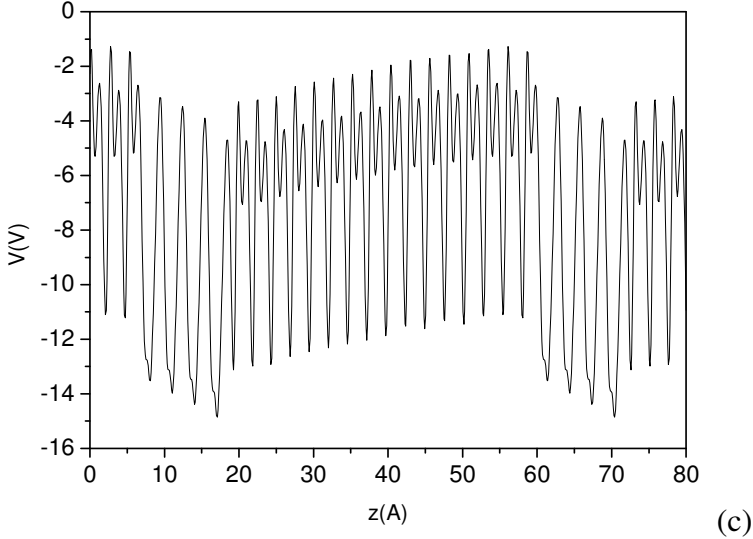

(c)

*Fig. 2 Average potential profiles, for the wells of eight (a), six (b) and four(c) In-N ALs, and for the barrier of sixteen Ga-N ALs, obtained from DFT SIESTA code.*

From these diagrams it follows that the electric field in the well is different for various well widths. Also the potential difference at both edges of the well is not the same. Generally these differences are huge, attaining more than 1.5 V in QW. It is worth noting that the field change is not related to the screening, absent in our model, it stems from the PBC conditions expressed in Eq. 1. The numerical data containing comparison of the fields obtained from polarization analysis and DFT simulations is presented in Table 1. It has to be stressed out that in addition to the field a considerable potential jumps are observed. This is a completely new effect which stems from the existence of dipole layers on both GaN-InN and InN-GaN interfaces (heterojunctions).

*Table 1. Elastic deformations, electric potential jumps (in V) electric fields (in MV/cm) obtained from polarization analysis and DFT simulations for 16 AL GaN barrier and different thickness of InN QWs.*

| ALs GaN-16 | $\epsilon_1$ (InN) | $\epsilon_3$ (InN) | $\epsilon_1$ (GaN) | $\epsilon_3$ (GaN) | $V_1$ | $V_2$ | $E_{3,InN}$ (pol) | $E_{3,InN}$ (DFT) | $E_{3,GaN}$ (pol) | $E_{3,GaN}$ (DFT) |
|---|---|---|---|---|---|---|---|---|---|---|
| InN-8 | -0.073 | 0.033 | 0.028 | -0.018 | -1.64 | 1.33 | -7.04 | -10.17 | 4.69 | 6.43 |
| InN-6 | -0.058 | 0.036 | 0.045 | -0.036 | -1.49 | 1.23 | -8.20 | -11.17 | 4.04 | 5.28 |
| InN-4 | -0.082 | 0.050 | 0.019 | -0.056 | -1.88 | 1.28 | -7.31 | -17.31 | 3.43 | 6.18 |

As it follows from the above data, the potential jumps are considerable, exceeding 1 V in all cases, which makes the InN well shallower. The localization of the carriers in QW is less efficient and that allows for their easier escape from the wells.

Another interesting aspect is the field dependence on the barrier thickness. It is frequently argued that the fields in the wells do not depend on the thickness of the barriers. The DFT data, obtained for six AL thick InN well and for different GaN barrier thickness, presented in Fig. 3, show large variation, in contradiction to standard arguments .

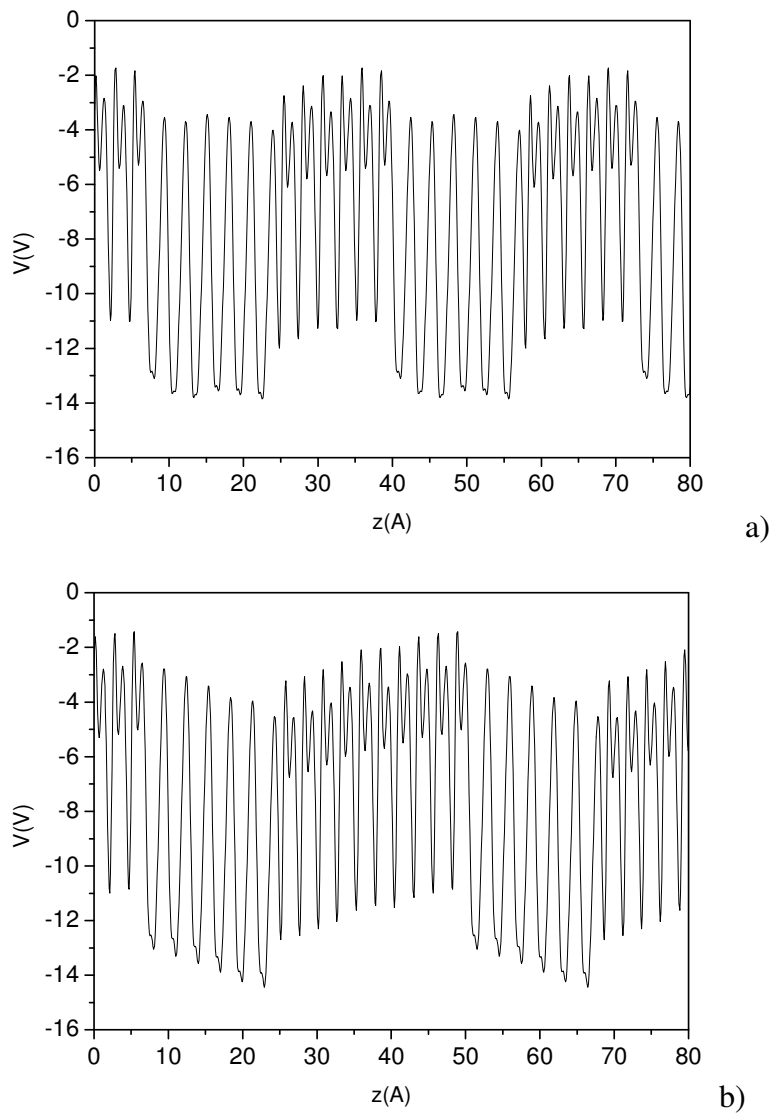

a)

b)

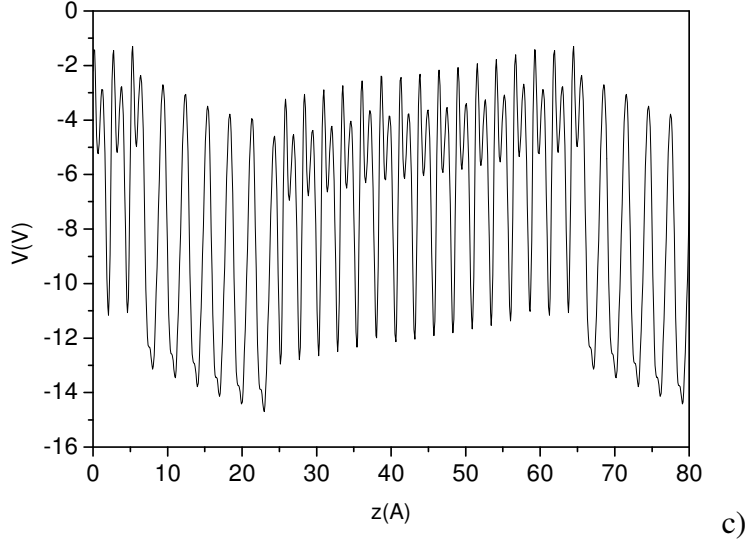

c)

*Fig. 3 Average potential profiles, for the wells of six In-N ALs, and the barriers of six(a), ten(b) and sixteen(c) Ga-N ALs, obtained from DFT SIESTA simulation.*

The fields in the well depend on the barrier thickness (see Table 2). It is also evident that for the lowest thickness of the barrier, the field in the well is not uniform. This may be related to the additional charge located at the interfaces. For the larger barrier thickness, the field is virtually constant (the potential profile is linear). The magnitude of the field is large, the difference of potential is about 1.4 V, the largest value for the thickest barrier, which confirms analysis given above. The detailed comparison of the polarization and DFT results, presented in Table 2 indicates on reasonable agreement of these two results.

*Table 2. Elastic deformations, electric potential jumps (in V) electric fields (in MV/cm) obtained from polarization analysis and DFT simulations for 6 AL InN QWs and different thickness of GaN barriers.*

| ALs    | $\epsilon_1$ (InN) | $\epsilon_3$ (InN) | $\epsilon_1$ (GaN) | $\epsilon_3$ (GaN) | $V_1$ | $V_2$ | $E_{3,InN}$ (pol) | $E_{3,InN}$ (DFT) | $E_{3,GaN}$ (pol) | $E_{3,GaN}$ (DFT) |
|--------|--------------------|--------------------|--------------------|--------------------|-------|-------|-------------------|-------------------|-------------------|-------------------|
| InN-6  |                    |                    |                    |                    |       |       |                   |                   |                   |                   |

| | | | | | | | | | |
|---|---|---|---|---|---|---|---|---|---|
| GaN-6 | -0.051 | 0.018 | 0.053 | -0.059 | -2.01 | 0.88 | -6.38 | -2.24 | 13.28 | 9.15 |
| GaN-10 | -0.058 | 0.028 | 0.045 | -0.029 | -1.21 | 0.64 | -6.60 | -11.11 | 5.49 | 7.85 |
| GaN-16 | -0.058 | 0.036 | 0.045 | -0.036 | -1.49 | 1.23 | -8.20 | -11.17 | 4.02 | 5.27 |

Another interesting aspect is direct verification of the presence of the dipole layer. It is worth noting that, according to Eq. 4, in absence of the interface dipole layer contribution (i.e. $\Delta V_1 = \Delta V_2 = 0$), the field in the wells and in the barriers depends on the InN-GaN thickness ratio. Thus a straightforward approach is to use the same thickness of both layers, i.e for $z_{InN}/z_{GaN} = 1$. Such data are presented in Fig.4.

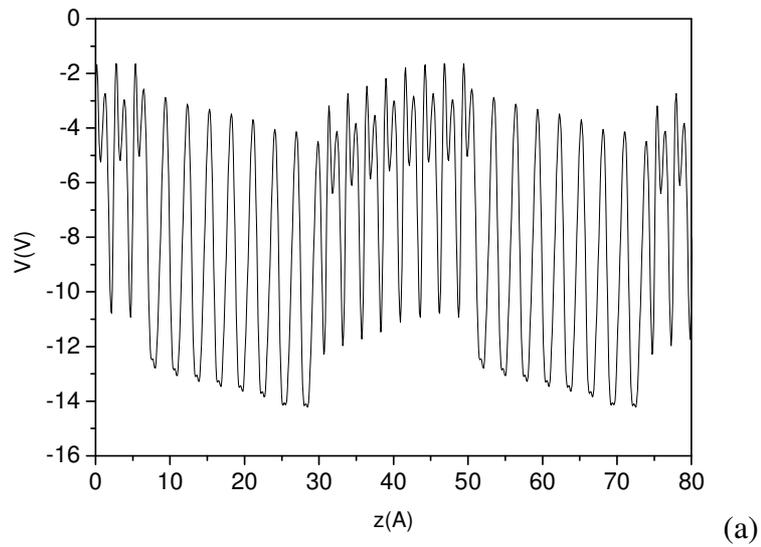

(a)

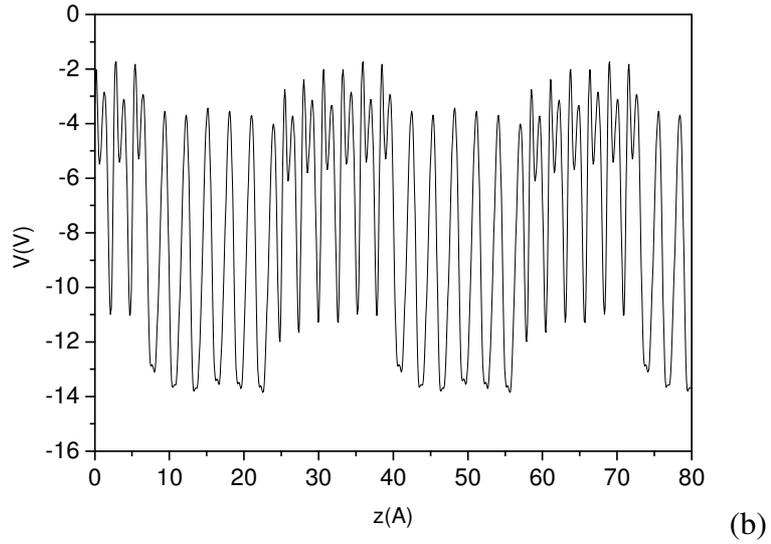

(b)

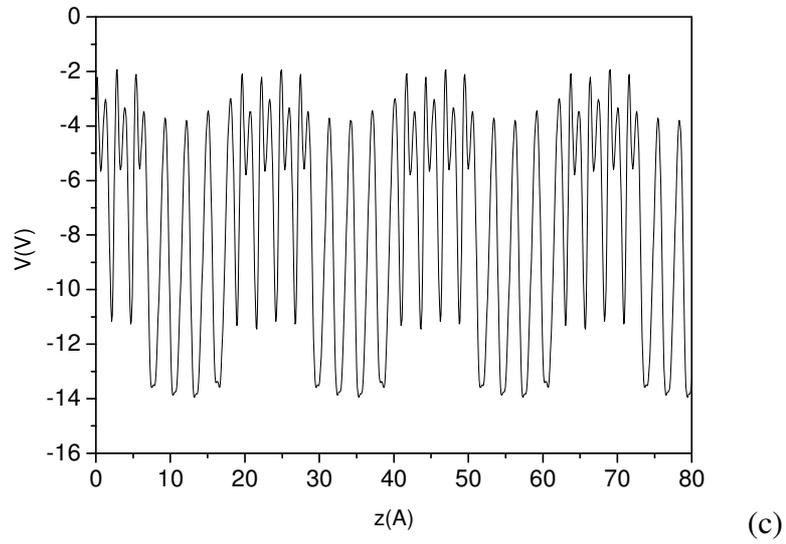

(c)

*Fig. 4 Electric potential profiles, drawn along the channeling path in GaN-InN MQW system, for the identical number of ALs in both InN wells and GaN barriers eight(a), six(b) and four(c), obtained from DFT SIESTA code.*

From the results shown in Fig.4 it follows that the fields are evidently different. It has to be added that the 4 AL case does not recover the linear distribution of the potential, which is most likely due to the overlap of the charges and dipoles from both interfaces. These simulation results and polarization analysis are summarized in Table 3.

*Table 3. Elastic deformations, electric potential jumps (in V) electric fields (in MV/cm) obtained from polarization analysis and DFT simulations for identical number of atomic layers (ALs) in InN QW and GaN barrier.*

| ALs | $\epsilon_1$ (InN) | $\epsilon_3$ (InN) | $\epsilon_1$ (GaN) | $\epsilon_3$ (GaN) | $V_1$ | $V_2$ | $E_{3,InN}$ (pol) | $E_{3,InN}$ (DFT) | $E_{3,GaN}$ (pol) | $E_{3,GaN}$ (DFT) |
|---|---|---|---|---|---|---|---|---|---|---|
| GaN-4 & InN-4 | -0.051 | 0.019 | 0.053 | -0.044 | -1.79 | 0.77 | -5.41 | 4.70 | 14.58 | 4.60 |
| GaN-6 & InN-6 | -0.051 | 0.019 | 0.053 | -0.055 | -2.01 | 0.88 | -6.38 | -2.24 | 13.29 | 9.15 |
| GaN-8 & InN-8 | -0.050 | 0.031 | 0.054 | -0.038 | -1.25 | 1.34 | -6.81 | -7.86 | 6.61 | 7.70 |

These data show that the DFT and polarization analysis results are in reasonable agreement. The only exception is the 4 AL case when the continuous approximation is not applicable. Altogether, the above results confirm also the validity of the energy minimization approach, in conjunction with the polarization results of Bernardini et al. and also prove applicability of DFT direct simulations to quantum well systems. [3].

The developed approach opens the route to simulation electric fields in two-dimensional quantum well structures directly, from ab initio simulations. The polarization analysis indicates that the fields in the well depend on the thickness of both barriers and wells due to PBC. The obtained fields are very high, that attains the energy difference above 1.5 eV in six ALs thick InN quantum well, influencing carrier distribution within the wells and negatively affecting efficiency of InGaN/GaN based LDs and LEDs.

A new, most important contribution is identification of dipole layer at the GaN-InN interfaces, i.e. completely different electric character of polar InN-GaN heterostructures. The effect considerably changes the depth of InN quantum wells. Since the potential jumps may exceed 1 V that negatively affects localization of the carriers in InN QWs.

**Acknowledgement**

The research was supported by the European Union within European Regional Development Fund, through grant Innovative Economy (POIG.01.01.02-00-008/08). The calculations were made using computing facilities of Interdisciplinary Centre for Mathematical and Computational Modelling of Warsaw University (ICM UW).